%% file: CHL-ICTR-PHE12-arXiv-2.tex
\documentclass{article}
\usepackage{amssymb}
\usepackage{graphicx}
\begin{document}

\include{00README.XXX}

\title{Submillimeter nuclear medical imaging with a Compton Camera using triple coincidences of collinear $\beta^+$-annihilation photons and $\gamma$ rays}

\author{C. Lang$^1$, D. Habs$^1$, P.G. Thirolf$^1$, A. Zoglauer$^2$}
\date{}

\maketitle

\begin{center}
\footnotesize
$^1$Fakult\"at f\"ur Physik, Ludwig-Maximilians Universit\"at M\"unchen, Garching, Germany
\newline
$^2$Space Science Laboratory, University of California, Berkeley, CA, USA
\end{center}
\vspace{0.5cm}

PET cameras had their breakthrough as an imaging instrument for clinical application studies due to the large variety of radioactively labeled tracer molecules [1]. These tracers carry an e$^+$ emitting isotope to e.g. a tumor. Annihilation of the positron into two back-to-back 511 keV photons allows to restrict the source origin in 2 dimensions onto a line of response (LOR).

Superimposing the LORs of different decay events locates the source distribution of the emitter in 3D. Modern PET systems reach a spatial resolution of 3-10 mm. A disadvantage of this technique is the diffusion of the positron before its decay with a typical range of ca. 1 mm (depending on its energy). This motion and Compton scattering of the 511 keV photons within the patient limit the performance of PET. 
We present a nuclear medical imaging technique, able to reach submillimeter spatial resolution in 3 dimensions with a reduced activity application compared to conventional PET. This 'gamma-PET' technique draws on specific e$^+$ sources simultaneously emitting an additional photon with the $\beta^+$ decay. Exploiting the triple coincidence between the positron annihilation and the third photon, it is possible to separate the reconstructed 'true' events from background [2,3]. Therefore the spatial uncertainty introduced by the motion of the e$^+$ or by Compton scattering within the patient can be strongly reduced in the direction normal to the annihilation. Especially $^{44}$Sc is of interest, which $\beta^+$ decays into $^{44}$Ca, emitting an 1157 keV photon. With its halflife of 3.9 h, $^{44}$Sc has to be produced from a $^{44}$Ti generator (t$_{1/2}$ = 60.4 a) [4].
Presently this cannot be performed in clinically relevant quantities, however, this may change with the soon expected availability of highly brilliant gamma beams [2]. $^{44}$Sc has already been applied in patient studies [5]. 
We combined the Compton camera technique, i.e. the measurement of photon energies and positions of Compton scattering interactions, with a PET camera defining the LOR. Due to the Compton kinematics and subsequent photon absorption, a Compton camera can reconstruct the origin of a primary photon on the surface of a 'Compton cone' [6].
Superimposing such cones from different events reduces the reconstructed source distribution in 3D to the few-millimeter range. The gamma-PET technique is different, it will intersect the Compton cone with the LOR from the same $\beta^+$ annihilation/gamma coincidence event, thus allowing to reconstruct the source position in 3D from single events.

In order to test the feasibility of this technique, Monte-Carlo simulations and image reconstruction has been performed using the 'Medium Energy Gamma-Ray Astronomy' library MEGAlib [7]. MEGAlib is a software framework designed to simulate and analyze data from Compton cameras. For the requirements of the gamma-PET technique we modified MEGAlib to realize an event reconstruction from the intersection of the Compton cone with the LOR.
The simulated geometry consists of four Compton camera modules in a quadratic arrangement surrounding an H$_2$O sphere of 6 cm diameter in a distance of 3.5 cm from the sphere center. Two of them serve for detecting the annihilation photons defining the LOR, the others allow for detecting the third photon, enabling to reconstruct the Compton cone. Inside the sphere two $^{22}$Na point-like, non-medical test sources were placed at a distance of 0.4 mm. Each camera module consists of a LaBr$_3$ scintillator crystal (50x50x30 mm$^3$) read out by a 2D-segmented photomultiplier with 64 pixels (6x6 mm$^2$ each) and a double-sided silicon strip detector with 128 strips on each side, an active area of 50x50 mm$^2$ and an optimized thickness of 2 mm.

Figure 1 shows the source distribution after 60 iterations of the reconstruction algorithm. 
\begin{figure}
\includegraphics[scale=0.4]{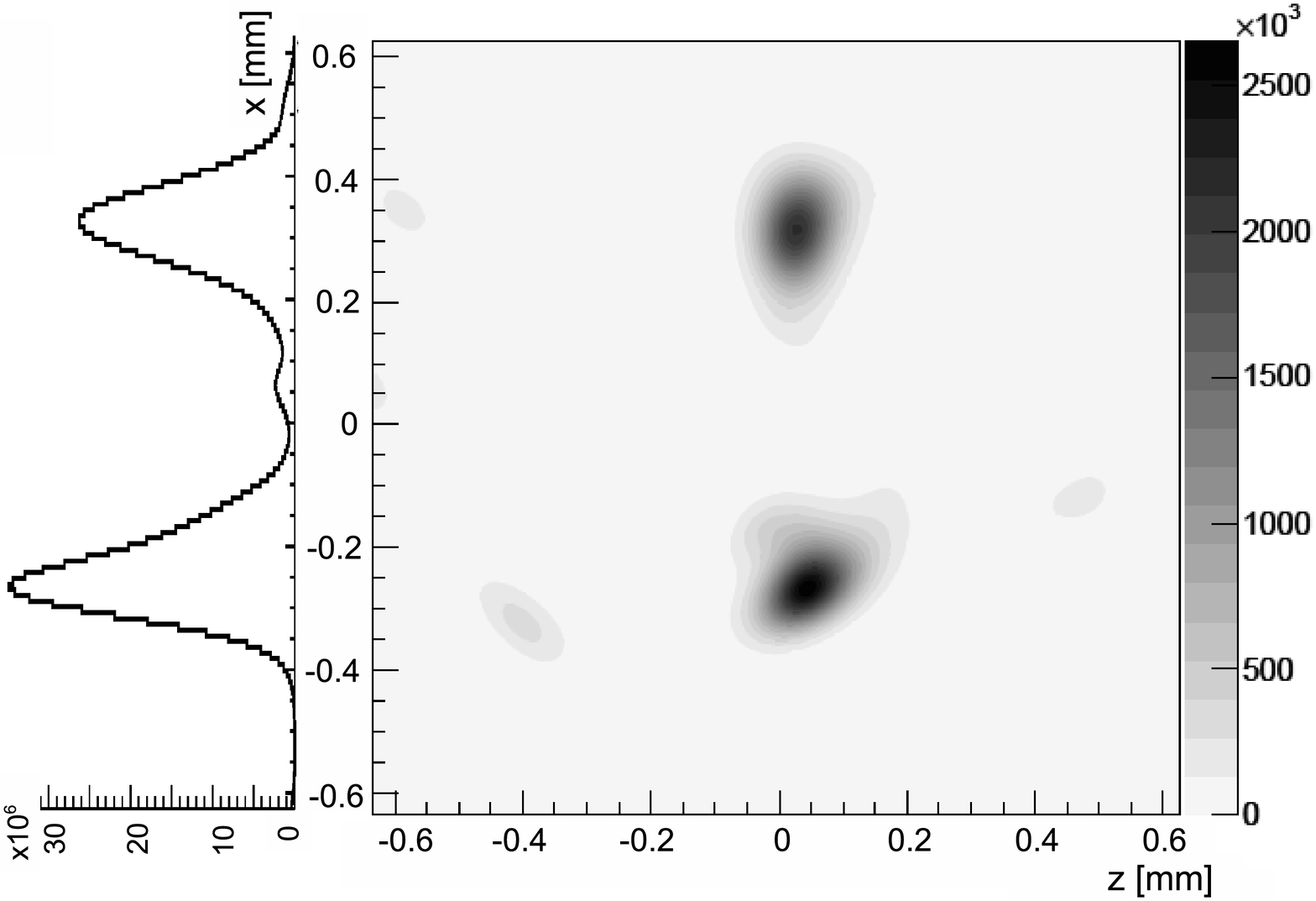}
\caption{Reconstructed source distribution of two non-medical, long-lived $^{22}$Na point-like test sources placed in a water sphere (6 cm diameter) with a distance of 0.4 mm between them. The reconstruction was based on triple coincidences between the two annihilation photons and the third 1275 keV gamma ray after 60 iterations of the reconstruction algorithm. (Left) ungated projection on the x axis.}
\end{figure}   

The two $^{22}$Na sources can clearly be resolved. The spatial resolution amounts to 0.2 mm (FWHM) in each direction, surpassing the performance of
conventional PET by about an order of magnitude. The simulated detector geometry exhibits a coincidence detection efficiency of 1.92e-7 per $^{22}$Na decay. In 25 $\%$ of these events the intersection between the LOR and the Compton cone could be identified. Starting with only 0.7 MBq of source activity (ca. 200-500 times less compared to conventional PET), about 20 intersections (sufficient for source reconstruction) can be identified after an exposure time of 450 s.
\newline
\newline
This work was supported by the DFG Cluster of Excellence MAP (Munich-Centre for Advanced photonics).
\newline
\newline
[1] P. Suetens, Fundamentals of Medical Imaging, Camb. Univ. Press (2002). 
\newline
[2] D. Habs and U. Koester, Appl. Phys. B 103, 501 (2011).
\newline
[3] C. Grignon et al., Nucl. Instr. Meth. A 571, 142 (2007).
\newline
[4] M. Pruszynski et al., Appl. Rad. Isot. 68, 1636 (2010).
\newline
[5] F. Roesch and R.P. Baum, Dalton Trans. 40, 6104 (2011).
\newline
[6] G. Kanbach et al., Nucl. Instr. Meth. A 541, 310 (2005).
\newline
[7] A. Zoglauer et al., New Astron. Rev. 50, 629 (2006).

\end{document}

%% file: 00README.XXX
CHL-ICTR-PHE12-arXiv-2.dvi keepcomments